\documentclass[twocolumn,aps,prl,preprintnumbers]{revtex4}
\usepackage{bbm}
\usepackage{amsmath}
\usepackage{amssymb}
\usepackage{graphicx}
\usepackage{mathrsfs}
\usepackage{amsfonts}
\usepackage{amsthm}
\usepackage{color}
\usepackage{txfonts}
\usepackage[colorlinks=true,citecolor=blue,linkcolor=blue,urlcolor=blue,anchorcolor=blue]{hyperref}%
\usepackage{pifont}
\hypersetup{colorlinks=true,citecolor=blue,linkcolor=blue,urlcolor=blue}
\providecommand{\U}[1]{\protect\rule{.1in}{.1in}}
\makeatletter
\@ifundefined{textcolor}{}
{
\definecolor{BLACK}{gray}{0}
\definecolor{WHITE}{gray}{1}
\definecolor{RED}{rgb}{1,0,0}
\definecolor{GREEN}{rgb}{0,1,0}
\definecolor{BLUE}{rgb}{0,0,1}
\definecolor{CYAN}{cmyk}{1,0,0,0}
\definecolor{MAGENTA}{cmyk}{0,1,0,0}
\definecolor{YELLOW}{cmyk}{0,0,1,0}
}

\makeatother
\begin{document}
\title{Stabilization and dynamics of magnetic antivortices in a nanodisk with anisotropic Dzyaloshinskii-Moriya interaction}
\author{Xin Hu}
\author{X.S. Wang}
\email[Corresponding author: ]{justicewxs@hnu.edu.cn}
\author{Zhenyu Wang}
\email[Corresponding author: ]{vcwang@hnu.edu.cn}

\affiliation{School of Physics and Electronics, Hunan University, Changsha 410082, China}

\begin{abstract}
We theoretically investigate the antivortex stabilized by anisotropic Dzyaloshinskii-Moriya interaction (DMI) in nanodisks. It is remarkably found that the antivortex remains stable even when the nanodisk radius is reduced to 15 nm, owing to the short-range nature of the DMI. We also investigate the antivortex dynamics under a static in-plane magnetic field, which shows that the displacement of the antivortex core depends on its vorticity and helicity, providing a fundamental basic for distinguishing different vortex types. Additionally, spin-polarized currents can trigger a self-sustained gyration of the antivortex at low current densities, while inducing polarity switching at high current densities. Our findings offer valuable insights into the DMI role in stabilizing topological solitons and their potential applications in spin-torque nano-oscillators and magnetic memories.
\end{abstract}

\maketitle
\section{Introduction}\label{sec1}
The competition between magnetic interactions in magnets gives rise to various magnetic textures including magnetic domain wall \cite{Venkat2023}, vortex \cite{Guslienko2008}, skyrmion \cite{Nagaosa2013}, hopfion \cite{Wang2019}, and so on.
Thereinto, magnetic vortex is a common ground state in a circular nanodisk resulting from the competition between exchange and dipolar interactions \cite{Raabe2000}, which can be characterized by its polarity and chirality, which refer to the direction of vortex core ($p=+1$ for upward and $p=-1$ for downward) and the rotation of the in-plane magnetization around the vortex core ($c=+1$ for counterclockwise and $c=-1$ for clockwise), respectively. Such a binary feature of magnetic vortex makes it being a good candidate for future data storage devices, such as vortex-based random access memory \cite{Bohlens2008,Kim2008}. Furthermore, magnetic vortex exhibits a variety of interesting dynamic properties driven by magnetic fields or spin-polarized currents, which have potential applications in spin-torque nano-oscillators \cite{Petit2012,Devolder2019}, neuromorphic computing \cite{Wu2024,Li2024}, and biomedicine \cite{Kim2010,Yang2015}.

Magnetic antivortex is a topological counterpart of magnetic vortex, exhibiting fascinating dynamics and offering potential applications in data storage \cite{Wang2007,Drews2009,Kamionka2010}. However, magnetic antivortex has received considerably less attention, which is mainly because of the difficulty in forming it. From the periphery to the antivortex core, the magnetization sweeps inwards along two opposite directions and outwards along the two perpendicular directions [see Fig. \ref{fig2}(a)], which would generate magnetic charges resulting in additional magnetostatic energy. To stabilize the antivortex, some specially shaped nanostructures such as asteroid \cite{Gliga2008,Xing2008}, $\infty$-shaped \cite{Kamionka2011}, $\varphi$-shaped \cite{Pues2012}, cross-like \cite{Mironov2010}, and pound-key-like patterns \cite{Haldar2013}, are required to design. But the fabrication of such complex nanopatterns hinders applications of magnetic antivortex. Thus, the generation of a stable isolated antivortex in the simplest geometry (e.g., a circular nanodisk) is very much desired.

\begin{figure}
  \centering
  \includegraphics[width=0.5\textwidth]{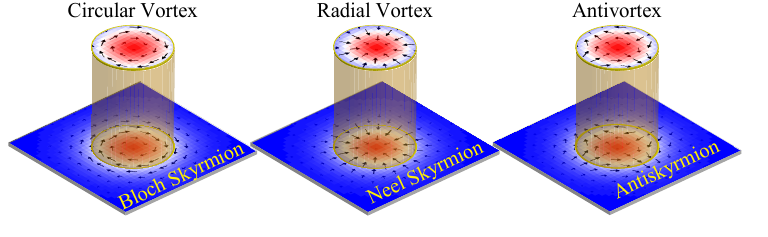}\\
  \caption{Schematic images of magnetization distributions for three different types of skyrmions (bottom panel) and the corresponding vortex types confined in the circular nanodisk (upper panel).}\label{fig1}
\end{figure}

The Dzyaloshinskii-Moriya interaction (DMI) is an antisymmetric exchange coupling \cite{Dzyaloshinsky1958,Moriya1960}, which has recently drawn extensive research interest due to its fundamental role in stabilizing skyrmions. Skyrmions stabilized by different DMI types have different spin configurations. For instance, the bulk DMI in chiral magnets favors the formation of the Bloch-type skyrmion \cite{Muhlbauer2009,Yu2010}, while the interfacial DMI in magnetic thin film covered by a heavy metal prefers the N{\'e}el-type skyrmion \cite{Heinze2011,Sampaio2013}. Most recently, it was demonstrated that the anisotropic DMI with opposite signs along two perpendicular directions can stabilize antiskyrmions \cite{Hoffmann2017,Huang2017,Camosi2018}. Moreover, the DMI effect on magnetic vortex has also been widely investigated. These works show that the DMI can lift the degeneracy of magnetic vortex with opposite chiralities and modify the size of vortex core \cite{Butenko2009,Luo2014,Wang2015}. Besides the static configuration, the DMI also has a great influence on the gyration \cite{Luo2015,Liu2016, Wei2023}, internal modes \cite{Mruczkiewicz2017,Mruczkiewicz2018,Flores2020}, and polarity switching \cite{Wang2020,Li2022} of magnetic vortex. Moreover, the radial vortex stabilized by the interfacial DMI has been theoretically demonstrated \cite{Siracusano2016} and observed in experiments \cite{Karakas2018}. In fact, circular and radial vortices in confined nanodisks can be viewed as partial Bloch-type and N{\'e}el-type skyrmions, which can be obtained by cutting along the skyrmion wall ($m_z=0$) from the magnetic thin film, as shown in Fig. \ref{fig1}. Similarly, the antivortex can be viewed as a partial antiskyrmion, which inspires us to explore the possibility of the antivortex stabilized by the anisotropy DMI in a circular nanodisk.

In this work, we investigate the stability of the antivortex in circular nanodisks with the anisotropic DMI. It is found that the radius of the nanodisk hosting the antivortex as a ground state can scale down to 15 nm, much smaller than the size of the nanodisk stabilizing the circular vortex induced by the competition between exchange and dipolar interactions. The antivortex profile is analytically calculated and numerically confirmed by micromagnetic simulations. By quasi-statically decreasing the out-of-plane magnetic field from a perpendicular single-domain state, the antivortex can be reliably generated. When subjected to an in-plane magnetic field, the antivortex exhibits movement in a different direction compared to circular and radial vortices. Furthermore, we demonstrate that spin-polarized currents can drive the gyration and polarity switching of the antivortex, a crucial aspect for applications in spin-torque nano-oscillators and antivortex-based random access memories.

The paper is organized as follows. In Sec. \ref{sec2}, we present the analytical model describing the antivortex profile. Section \ref{sec3} gives the numerical results to verify theoretical analysis. We also discuss the stability of the antivortex as a function of the disk size and DMI strength, the antivortex nucleation, the dynamics of the antivortex under in-plane magnetic fields and spin-polarized currents. Discussion and conclusion are drawn in Sec. \ref{sec4}.

\section{Theoretical model}\label{sec2}
We consider a magnetic nanodisk with the radius $R$ and thickness $t$, whose energy is written as
\begin{equation}\label{eq_Energy}
  E_{\mathrm{tot}}=t\int\Big(A\nabla^{2}\mathbf{m}+K_{m}m_{z}^{2}+w_{D}\Big)d\mathbf{r},
\end{equation}
where $\mathbf{m}=\mathbf{M}/M_s$ is the normalized magnetization with the saturated magnetization $M_s$, and $A$ is the exchange stiffness. The demagnetization energy is approximated by a surface anisotropy with the effective anisotropy constant $K_m=\mu_0 M_s^2/2$. $w_D$ is the DMI energy density, which has the form of \cite{Camosi2018}
\begin{equation}\label{eq_DMI_energy}
  w_D=D_x\Bigg(m_z\frac{\partial m_x}{\partial x}-m_x\frac{\partial m_z}{\partial x}\Bigg)+D_y\Bigg(m_z\frac{\partial m_y}{\partial y}-m_y\frac{\partial m_z}{\partial y}\Bigg),
\end{equation}
where $D_x$ and $D_y$ are the DMI constants along the $\mathbf{x}$ and $\mathbf{y}$ directions. Here, we focus on the anisotropic DMI with the equal magnitude but with opposite signs ($D_x=-D_y=D$). We express the magnetization vector in spherical coordinate $\mathbf{m}=\{\sin\theta\cos\varphi, \sin\theta\sin\varphi, \cos\theta\}$, as shown in Fig. \ref{fig2}(b).

In the polar coordinates $\mathbf{r}=(\rho, \phi)$ originated at the disk center, we assume that the radial profile of the polar angle $\theta$ depends on $\rho$ only, which means the radial profiles along any ray from the center are the same. The azimuthal angle $\varphi$ of magnetization can be expressed by $\varphi(\phi)=\nu\phi+\gamma$ with the vorticity (or called winding number) $\nu$ and the helicity $\gamma$. The circular and radial vortices have the equal vorticity ($\nu=1$) but different helicites ($\gamma=\pm\pi/2$ for circular vortex and $\gamma=0,\pi$ for radial vortex).
For the antivortex discussed here [see Fig. \ref{fig2}(a)], the vorticity and helicity are $\nu=-1$ and $\gamma=\pi$, respectively. Thus, antivortex profile $\mathbf{m}(\mathbf{r})$ can be written in a variable-separated form as  $\theta=\theta(\rho)$ and $\varphi=\varphi(\phi)$. The system energy can be written as
\begin{equation}\label{eq_Energy_Polar}
  \begin{split}
     E_{\mathrm{tot}} &= 2\pi t \int \Bigg\{A\Bigg[\Bigg(\frac{d\theta}{d\rho}\Bigg)^{2}+\frac{\sin^{2}\theta}{\rho^{2}}\Bigg]+K_m\cos^{2}\theta \\
       & -D\Bigg(\frac{d\theta}{d\rho}+\frac{\cos\theta\sin\theta}{\rho}\Bigg)\Bigg\}\rho d\rho.
   \end{split}
\end{equation}
Minimizing the energy function Eq. (\ref{eq_Energy_Polar}) with the Euler-Lagrange method leads to the equation of $\theta(\rho)$:
\begin{equation}\label{eq_theta_pho}
  A\Bigg(\frac{d^{2}\theta}{d\rho^{2}}+\frac{1}{\rho}\frac{d\theta}{d\rho}-\frac{\sin\theta\cos\theta}{\rho^{2}}\Bigg)
  +K_m\sin\theta\cos\theta-\frac{D}{\rho}\sin^{2}\theta=0,
\end{equation}
with the boundary condition \cite{Rohart2013}
\begin{equation}\label{eq_BC}
\theta|_{\rho=0}=0, \quad
\frac{d\theta}{d\rho}\Big|_{\rho=R}=\frac{D}{2A}.
\end{equation}
The radial profile of the antivortex $\theta(\rho)$ can be obtained by numerically solving the auxiliary Cauchy problem Eqs. (\ref{eq_theta_pho}) and (\ref{eq_BC}) using the Runge-Kutta method \cite{Butenko2009}.

\section{Numerical results}\label{sec3}
\subsection{Magnetization configuration of the antivortex}
\begin{figure}
  \centering
  \includegraphics[width=0.5\textwidth]{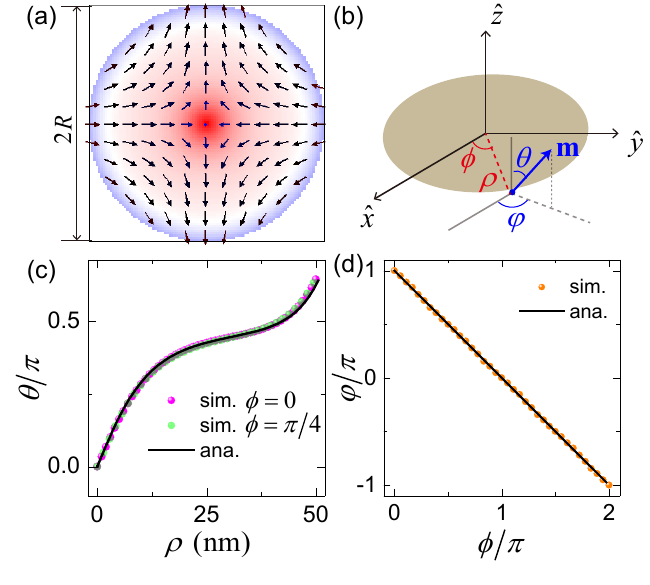}\\
  \caption{(a) Static configuration of the antivortex in a nanodisk with radius $R=50$ nm, $t=1$ nm, and $D=2.5$ $\mathrm{mJ/m^2}$. (b) Schematic of the polar and spherical coordinations. (c) The polar angle of the antivortex profile $\theta(\rho)$ along two different directions $\phi=0$ and $\pi/4$. (d) The azimuthal angle of the antivortex profile at $\rho=25$ nm. In (c) and (d), the color dots are simulation data and the black curves are analytical results.}\label{fig2}
\end{figure}

To verify the above theoretical analysis, we perform micromagnetic simulations using the Mumax3 code \cite{Vansteenkiste2014} including the anisotropic DMI. A ferromagnetic nanodisk with the radius of $R=50$ nm and thickness of $t=1$ nm is considered. Magnetic parameters of permalloy are used in simulations: $M_s=8.6\times10^{5}$ $\mathrm{A/m}$, $A_{ex}=13$ $\mathrm{pJ/m}$, and $\alpha=0.01$. The mesh size is set to $1\times1\times1$ $\mathrm{nm^3}$. For $D=2.5$ $\mathrm{mJ/m^{2}}$, the ground state of the nanodisk is an antivortex, shown in Fig. \ref{fig2}(a). The polar and azimuthal angles of the magnetization extracted from the simulation data are plotted in Figs. \ref{fig2}(c) and \ref{fig2}(d), which shows an excellent agreement with the analytical results. Note that the circular symmetry for spin rotations of the antivortex in Fig. \ref{fig2}(a) is broken, which results in the dipolar energy at the disk edge being anisotropic. To ascertain the influence of such anisotropic dipolar interaction on the antivortex profile, the magnetization configurations along different azimuthal angles ($\phi=0$, and $\pi/4$) are plotted in Fig. \ref{fig2}(c), showing a negligible difference. This justifies the separation of radial and azimuthal variables in Eq. \eqref{eq_Energy_Polar} and \eqref{eq_theta_pho}.

\subsection{Stabilization of the antivortex}
\begin{figure}
  \centering
  \includegraphics[width=0.45\textwidth]{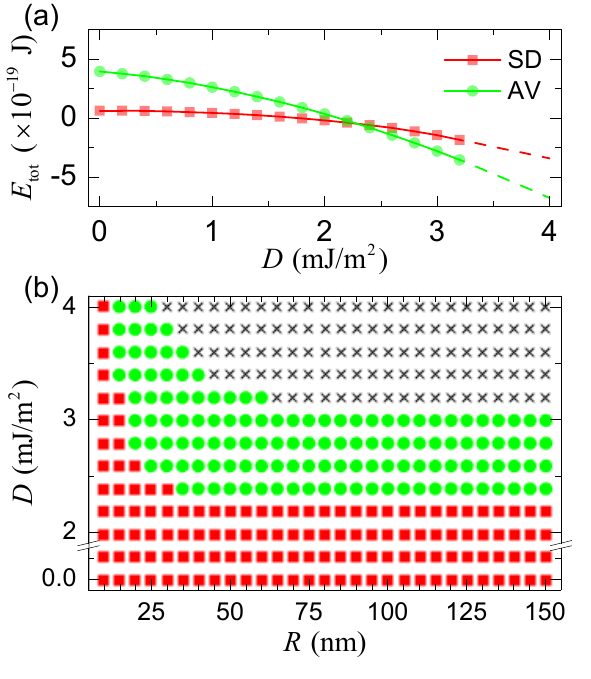}\\
  \caption{(a) Total energy of the nanodisk ($R=50$ nm and $t=1$ nm) with two different states [single domain (SD) and antivortex (AV) states] as a function of $D$. Dashed lines above $D=3.2$ $\mathrm{mJ/m^2}$ indicate that the corresponding states are unstable and relax to multi-domain states. (b) Phase diagram of the ground state of the nanodisk in ($R$, $D$) space. Red square, green dot, and black cross represent the single-domain, antivortex, and multi-domain states, respectively.}\label{fig3}
\end{figure}

We then explore the antivortex stability in the parameter space ($D$, $R$). Here, we fix the thickness of the nanodisk to $t=1$ nm. In the case of $D=0$, the ground state of the nanodisk ($R\leq150$ nm considered here) is a single-domain state. Thus, we choose two initial states (single-domain and antivortex) to relax and obtain the final equilibrium state. Figure \ref{fig3}(a) shows the total energy of the nanodisk with the two configurations, single-domain (SD) and antivortex (AV), as a function of $D$. For $0\leq D<2.3$ $\mathrm{mJ/m^{2}}$, the most stable state is the single-domain state. The antivortex becomes the ground state when $2.3<D\leq3.2$ $\mathrm{mJ/m^{2}}$, where the anisotropic DMI dominates over the exchange interaction. For $D>3.2$ $\mathrm{mJ/m^{2}}$,  multi-domain patterns arise to minimize the DMI energy (not shown here).

Next, we further investigate the stability of antivortex in nanodisks with different radii. Phase diagram obtained from micromagnetic simulations is given in Fig. \ref{fig3}(b). One can see that the antivortex state can remain stable ranging from $R=150$ nm down to $R=35$ nm when the DMI is between 2.4 and 3 $\mathrm{mJ/m^{2}}$. With the increase of the DMI, the size of the antivortex disk can be further reduced. When $D=3.4$ $\mathrm{mJ/m^{2}}$, the radius of the nanodisk hosting the antivortex can be as low as $R=15$ nm. This result is different from the conventional circular vortex, where the minimal radius of the nanodisk required to stabilize the vortex is about 450 nm for $t=1$ nm \cite{Metlov2002,Metlov2008}. Such a size limitation of the vortex formation primarily arises from the long-range nature of dipolar interaction. In contrast, the anisotropic DMI is the antisymmetric exchange interaction between nearest neighboring magnetizations, thereby reducing the size restriction on the antivortex formation. This mechanism is not only effective in stabilizing the antivortex by anisotropic DMI, but can also be employed to reduce the size of nanodisks hosting circular and radial vortices via bulk and interfacial DMIs. It is noted that the radius of the vortex disk can be reduced to 50 nm by coupling to the antidot matrix \cite{Verba2018, Verba2020}. However, accurately fabricating complex nanostructures in the nanoscale range poses a challenge in experiments. Therefore, the method presented in our work may be more suitable for miniaturizing vortex disks, potentially advancing the applications of vortices in magnetic recording.

\subsection{Nucleation of the antivortex}
\begin{figure}
  \centering
  \includegraphics[width=0.5\textwidth]{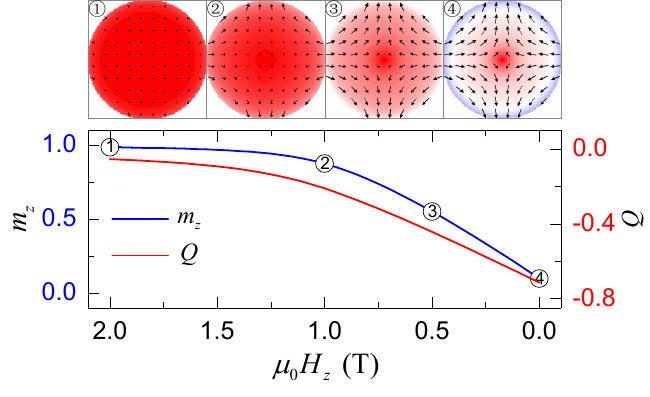}\\
  \caption{The z-component magnetization $m_z$ and the topological charge $Q$ of the nanodisk with $R=50$ nm and $D=2.5$ $\mathrm{mJ/m^2}$ under the perpendicular field $H_z$. The upper panel is the snapshots of the spatial distribution of the magnetization in the process of the antivortex nucleation.}\label{fig4}
\end{figure}

Because of the energy barrier between various magnetization configurations, achieving the antivortex state is not always feasible in the antivortex ground state region, and other metastable states (single-domain and multi-domain states) may arise instead. For practical applications, reliable generation of the antivortex is highly desired. A simple method is relaxing from an out-of-plane saturation state by quasi-statically decreasing the external field to zero, as has been done for the nucleation of radial vortex in Ref. \cite{Siracusano2016}. As the perpendicular field is gradually decreased, the magnetization tilting at the disk edge, which is caused by the anisotropic DMI, spreads towards the disk center, ultimately forming a stable antivortex at remanence, as depicted in Fig. \ref{fig4}. The topological charge is also calculated by using the formula $Q=(1/4\pi)\iint\mathbf{m}\cdot(\partial_{x}\mathbf{m}\times\partial_{y}\mathbf{m})dxdy$, as plotted in the red curve in Fig. \ref{fig4}. One can see that the topological charge of the antivortex is not $Q=-0.5$ but $Q=-0.7$, which is mainly because the magnetization at the disk boundary tilts to $-\hat{z}$ direction caused by the DMI.

\subsection{Response to a static in-plane external field}
\begin{figure}
  \centering
  \includegraphics[width=0.5\textwidth]{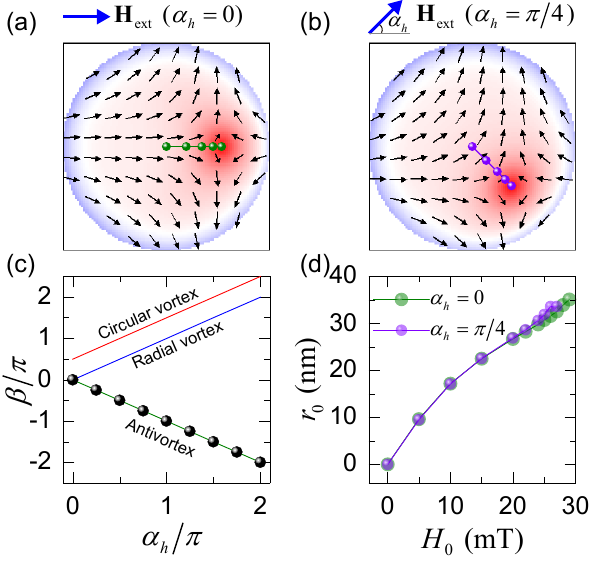}\\
  \caption{Trajectories of the vortex core under the in-plane field with two different directions (a) $\alpha_h=0$ and (b) $\alpha_h=\pi/4$. (c) The dependence of the shift direction of the vortex core on the field direction, the black dots are simulation results, and the solid lines are the analytical formula Eq. (\ref{eq_vcshift}). (d) The shift length $r_0$ as a function of the field amplitude $H_0$ for $\alpha_h=0$ and $\pi/4$.}\label{fig5}
\end{figure}

Different vortex configurations result in different responses to the external field. To describe the vortex response to the external field, we denote here that the shift of the vortex core is $\mathbf{r}_c=(r_0\cos\beta,r_0\sin\beta)$ under an
in-plane magnetic field $\mathbf{H}_\mathrm{ext}=(H_0\cos\alpha_h,H_0\sin\alpha_h)$. To minimize the Zeeman energy, the external field would enlarge the areas where the magnetization aligns with the field, while contracting the areas where the magnetization opposes the field. By analyzing the changes in the vortex structure, the shift direction (azimuthal angle) of the vortex core can be obtained,
\begin{equation}\label{eq_vcshift}
  \beta=\nu\alpha_h+\pi-\gamma.
\end{equation}
This result suggests that the response of the vortex to the in-plane field depends on both its vorticity ($\nu$) and helicity ($\gamma$).
For the circular and radial vortices, $\beta$ is equal to $\alpha_h+\pi/2$ and $\alpha_h$, indicating that the vortex core moves along the direction perpendicular and parallel to the field, respectively \cite{Siracusano2016,Verba2020}. For the antivortex ($\beta=-\alpha_h$), the vortex core shifts along the field direction for $\alpha_h=0$, and moves perpendicular to the field direction for $\alpha_h=\pi/4$, as shown in Figs. \ref{fig5}(a) and \ref{fig5}(b). Therefore, the distinct displacements of the vortex core under an in-plane magnetic field for the three vortex types can serve as a ``fingerprint" to distinguish them in experimental measurements, as illustrated in Fig. \ref{fig5}(c).

Besides the shift direction ($\beta$), we also study the shift displacement ($r_0$) varies with the field amplitude ($H_0$), as shown in Fig. \ref{fig5}(d). As expected, the shift length increases with the field amplitude before the antivortex annihilation. It's worth noting that the shift length is the same for two different field directions when the vortex core is away from the disk edge. When the vortex core nears the disk edge, the shift length is slightly larger for $\alpha_h=\pi/4$ compared to $\alpha_h=0$. This difference mainly stems from the anisotropic demagnetization. The edge magnetic charge density  maximizes at $\phi=0$ and vanishes at $\phi=-\pi/4$. For the circular and radial vortices, the dipolar interaction at the disk edge is isotropic.

\subsection{Gyration and switching of the antivortex induced by spin-polarized currents}
\begin{figure}
  \centering
  \includegraphics[width=0.5\textwidth]{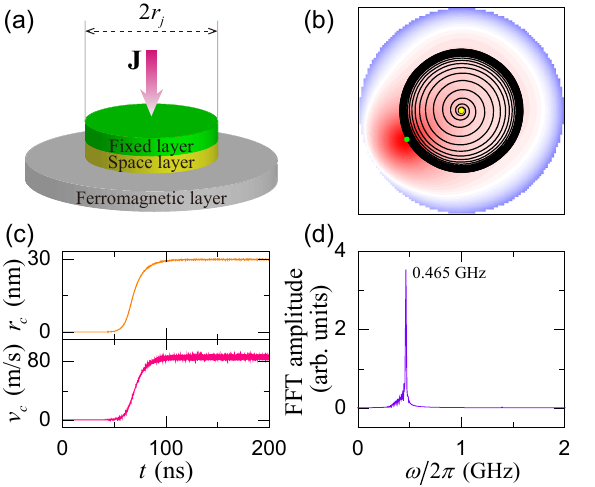}\\
  \caption{(a) Schematic diagram of the system consisting of a fixed layer, a space layer, and a ferromagnetic layer. The purple arrow represents the injected current. The radiuses of the nanocontact and ferromagnetic layer are $r_j=10$ nm and $R=50$ nm, respectively. (b) Trajectory of the antivortex core driven by the spin-polarized current. The yellow and green dots are the starting and ending points of the trajectory. (c) Temporal evolution of the gyrating orbit radius and velocity of the antivortex core. (d) The fast Fourier transform spectrum of the x-component of the antivortex core position.}\label{fig6}
\end{figure}

In the above sections, we have demonstrated the stabilization of antivortices and their field-induced dynamics. In this section, we continue to explore the dynamics of the antivortex when driven by spin-polarized currents. We consider the Slonczewski spin-transfer torque in simulations, which has the form
\begin{equation}\label{eq_STT}
  \mathbf{\tau}=\frac{\gamma \hbar P j_{0}}{2eM_{s}t}[\mathbf{m}\times(\mathbf{m}_{P}\times\mathbf{m})],
\end{equation}
where $\gamma$ is the gyromagnetic ratio, $\hbar$ is the reduced Plank constant, $e$ is the electron charge, $P$ is the spin polarization efficiency, $j_0$ is the current density, and $\mathbf{m}_P$ is the spin polarization vector. Here, we set $P=0.5$ and $\mathbf{m}_P=(0,0,-1)$. We also consider the Oersted field accompanying current flow in simulations, which can be calculated by Biot-Savart's law.
Figure \ref{fig6}(a) shows the schematic diagram of the system, where the nanocontact, for current injection, has radius $r_j$ and locates at the disk center. Below, the radius of the nanocontact is fixed at 10 nm.
At a low current density $j_0=2\times10^{10}$ $\mathrm{A/m^2}$, the antivortex core rotates clockwise around the disk center. As time goes by, it evolves into a steady self-oscillatory circular motion, reaching a maximum orbit radius of $r_c=30$ nm and moving at a speed of $v_c=87$ $\mathrm{m/s}$, as shown in Figs. \ref{fig6}(b) and \ref{fig6}(c). By the fast Fourier transform of the x-component or y-component of the position vector, we can obtain the gyrotropic frequency of the antivortex core $\omega_g/2\pi=0.465$ GHz [see Fig. \ref{fig6}(d)], which is consistent with $v_c/2\pi r_c=0.462$ GHz. The frequency of the antivortex core gyration can be modulated by changing the nanocontact size $r_j$ or current density $j_0$ \cite{Li2020}.

\begin{figure}
  \centering
  \includegraphics[width=0.5\textwidth]{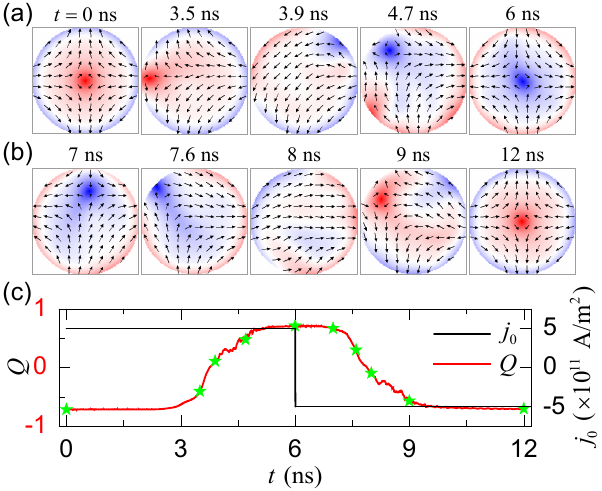}\\
  \caption{The antivortex core reversal driven by the spin-polarized current. (a) Snapshots of the antivortex core switching process for $j_0=5\times10^{11}$ $\mathrm{A/m^2}$. (b) Snapshots of the antivortex core ($p=-1$) switching process when the spin-polarized current is reversed ($j_0=-5\times10^{11}$ $\mathrm{A/m^2}$). (c) The topological charge $Q$ and spin-polarized current density $j_0$ as a function of time. Green stars represent the topological charges of the disk at various times in (a) and (b).}\label{fig7}
\end{figure}

We further increase the current density to $j_0=5\times10^{11}$ $\mathrm{A/m^2}$ for switching the antivortex core. Figure \ref{fig7}(a) shows the magnetization structure of the nanodisk during the antivortex core reversal.
Driven by the spin-polarized current, the antivortex with $p=1$ starts to gyrate with a continuously increasing orbit radius within 3.5 ns. At 3.9 ns, the antivortex core reaches the disk boundary and annihilates. As a result, the antivortex transforms into a quasi-uniform magnetization structure accompanied by edge solitons. These edge solitons rotate along the disk boundary and merge to form a downward antivortex core at 4.7 ns. Once the reversed antivortex core ($p=-1$) is formed, it undergoes a damped gyration and finally returns to the disk center at 6 ns. This entire process resembles the edge-soliton-mediated vortex core reversal described in Ref. \cite{Lee2011}.
The switching of the antivortex core can be further confirmed via the variation of the topological charge, which changes from $Q=-0.7$ at $t=0$ ns to $Q=0.7$ at $t=6$ ns [shown in Fig. \ref{fig7}(c)].
When the injection direction of the spin-polarized current is reversed ($j_0=-5\times10^{11}$ $\mathrm{A/m^2}$), the antivortex core with $p=-1$ reverses back to the $p=+1$ state, as illustrated in Fig. \ref{fig7}(b). The antivortex polarity switching by using spin-polarized current present here, provides a reliable and effective approach for information writing in magnetic random access memory devices.

It is reasonable to expect that varying the size of the nanocontact or the ferromagnetic disk may reveal new reversal mechanisms, similar to the vortex-antivortex mediated mechanism \cite{Waeyenberge2006} and breathing switching mode \cite{Ma2019}. Future studies on the antivortex polarity reversing can utilize other methods, including alternating magnetic field \cite{Waeyenberge2006,Curcic2008}, electric field \cite{Yu2020,Yu2021}, and strain \cite{Ostler2015,Zhu2021}, as well. Moreover, the quantitative analysis of the antivortex gyration and switching requires a theoretical derivation based on the Thiele equation, along with a systematical simulation, both of which are left for future study.

\section{Discussion and Conclusion}\label{sec4}
The anisotropic DMI discussed in this work has been both theoretically predicted and experimentally observed in magnetic materials with $D_{2d}$ or $C_{2v}$ symmetry \cite{Crepieux1998,Camosi2017,Hoffmann2017,Nayak2017}. Most recently, it has been demonstrated that the anisotropic DMI can be induced by applying different strains along two perpendicular directions \cite{Sun2023}, ensuring the viability of the antivortex induced by anisotropic DMI.

The anisotropic DMI-induced antivortex, as presented in this study, may inspire renewed interest and research into antivortices. One particularly intriguing aspect is the investigation of the internal modes of the antivortex.  Given that the rotational symmetry of the antivortex structure is broken, it might lead to a significant modification of azimuthal and radial spin-wave modes along with a notable mode coupling between them \cite{Cheng2024}. Second, the dipolar field generated by the antivortex is anisotropic at the disk boundary. Consequently, the energy transfer between dipolar-coupled antivortex disks is in dependence on the arrangement of the disk array \cite{Kim2012,Han2013}. In a two-dimensional lattice of antivortices, the bulk states are expected to exhibit a preferred direction of propagation, and there is a possibility for the emergence of novel topological states \cite{Li2019,Li202002}. Lastly, the gyration of the antivortex suggests its potential application in spin-torque nano-oscillators, similar to the circular and radial vortices \cite{Petit2012,Devolder2019,Li2020,Ma2020}. Moreover, the antivortex can also be utilized as a polarizer to induce a self-sustained gyration of antiskyrmions, analogous to the role of conventional vortices in skyrmion-based spin-torque nano-oscillators \cite{Garcia2016}.

In summary, we have shown that the anisotropic DMI can stabilize the antivortex in circular nanodisks with radius as small as 15 nm, extending well beyond the stability range of conventional vortices in an isolated nanodisk. The antivortex can be nucleated by decreasing the perpendicular magnetic field from an out-of-plane saturated state. Its movement under an in-plane magnetic field, depending on its vorticity and helicity, differs significantly from that of circular and radial vortices. Additionally, we have demonstrated the gyration and polarity switching of the antivortex triggered by spin-polarized currents. Our results enhance the fundamental understanding of anti-solitons in confined magnets and pave the way for applications in antivortex-based spintronic devices.

\section{Acknowledgment}
We thank R. Wang, C. Li, J. Sun, X. Liu, and Z. Zeng for helpful discussions. This work is supported by the Fundamental Research Funds for the Central Universities. Z.W. acknowledges the support the Natural Science Foundation of China (NSFC) (Grant No. 12204089). X.S.W. acknowledges the support from the NSFC (Grant No. 12174093).



\begin{thebibliography}{99}

\bibitem{Venkat2023} G. Venkat, D. A. Allwood, and T. J. Hayward, Magnetic Domain Walls: types, processes and applications, \href{https://doi.org/10.1088/1361-6463/ad0568}{J. Phys. D: Appl. Phys. \textbf{57}, 063001 (2023)}.
\bibitem{Guslienko2008} K. Yu. Guslienko, Magnetic Vortex State Stability, Reversal and Dynamics in Restricted Geometries, \href{https://doi.org/10.1166/jnn.2008.18305}{J. Nanosci. Nanotechnol. \textbf{8}, 2745 (2008)}.
\bibitem{Nagaosa2013} N. Nagaosa and Y. Tokura, Topological Properties and Dynamics of Magnetic Skyrmions, \href{https://doi.org/10.1038/nnano.2013.243}{Nat. Nanotechnol. \textbf{8}, 899 (2013)}.
\bibitem{Wang2019} X. S. Wang, A. Qaiumzadeh, and A. Brataas, Current-Driven Dynamics of Magnetic Hopfions, \href{https://doi.org/10.1103/PhysRevLett.123.147203}{Phys. Rev. Lett. \textbf{123}, 147203 (2019)}.
\bibitem{Raabe2000} J. Raabe, R. Pulwey, R. Sattler, T. Schweinb\"{o}ck, J. Zweck, and D. Weiss, Magnetization Pattern of Ferromagnetic Nanodisks, \href{https://doi.org/10.1063/1.1289216}{J. Appl. Phys. \textbf{88}, 4437 (2000)}.
\bibitem{Bohlens2008} S. Bohlens, B. Kr\"{u}ger, A. Drews, M. Bolte, G. Meier, and D. Pfannkuche, Current controlled random-access memory based on magnetic vortex handedness, \href{https://doi.org/10.1063/1.2998584}{Appl. Phys. Lett. \textbf{93}, 142508 (2008)}.
\bibitem{Kim2008} S.-K. Kim, K.-S. Lee, Y.-S. Choi, and Y.-S. Yu, Low-Power Selective Control of Ultrafast Vortex-Core Switching by Circularly Rotating Magnetic Fields: Circular-Rotational Eigenmodes, \href{https://doi.org/10.1109/TMAG.2008.2001539}{IEEE Trans. Magn. \textbf{44}, 3071 (2008)}.
\bibitem{Petit2012} S. Petit-Watelot, J.-V. Kim, A. Ruotolo, R. M. Otxoa, K. Bouzehouane, J. Grollier, A. Vansteenkiste, B. Van de Wiele, V. Cros, and T. Devolder, Commensurability and chaos in magnetic vortex oscillations, \href{https://doi.org/10.1038/nphys2362}{Nat. Phys. \textbf{8}, 682 (2012)}.
\bibitem{Devolder2019} T. Devolder, D. Rontani, S. Petit-Watelot, K. Bouzehouane, S. Andrieu, J. Letang, M. W. Yoo, J. P. Adam, C. Chappert, S. Girod, V. Cros, M. Sciamanna, and J. V. Kim, Chaos in Magnetic Nanocontact Vortex Oscillators, \href{https://doi.org/10.1103/PhysRevLett.123.147701}{Phys. Rev. Lett. \textbf{123}, 147701 (2019)}.
\bibitem{Wu2024} Y. Wu, Y. Luo, L. Zhang, S. Dai, B. Zhang, Y. Zhou, B. Fang, and Z. Zeng, Adjustable artificial neuron based on vortex magnetic tunnel junction, \href{https://doi.org/10.1063/5.0195602}{Appl. Phys. Lett. \textbf{124}, 122408 (2024)}.
\bibitem{Li2024} R. Li, Y. Rezaeiyan, T. B\"{o}hnert, A. Schulman, R. Ferreira, H. Farkhani, and F. Moradi, Temperature effect on a weighted vortex spin-torque nano-oscillator for neuromorphic computing, \href{https://doi.org/10.1038/s41598-024-60929-3}{Sci. Rep. \textbf{14}, 10043 (2024)}.
\bibitem{Kim2010} D.-H. Kim, E. A. Rozhkova, I. V. Ulasov, S. D. Bader, T. Rajh, M. S. Lesniak, and V. Novosad, Biofunctionalized Magnetic-Vortex Microdiscs for Targeted Cancer-Cell Destruction, \href{https://doi.org/10.1038/nmat2591}{Nat. Mater. \textbf{9}, 165 (2010)}.
\bibitem{Yang2015} Y. Yang, X. Liu, Y. Lv, T. S. Herng, X. Xu, W. Xia, T. Zhang, J. Fang, W. Xiao, and J. Ding, Orientation Mediated Enhancement on Magnetic Hyperthermia of $\mathrm{Fe_{3}O_{4}}$ Nanodisc, \href{https://doi.org/10.1002/adfm.201402764}{Adv. Funct. Mater. \textbf{25}, 812 (2015)}.
\bibitem{Wang2007} H. Wang and C. E. Campbell, Spin dynamics of a magnetic antivortex: Micromagnetic simulations, \href{https://doi.org/10.1103/PhysRevB.76.220407}{Phys. Rev. B \textbf{76}, 220407 (2007)}.
\bibitem{Drews2009} A. Drews, B. Kr\"{u}ger, G. Meier, S. Bohlens, L. Bocklage, T. Matsuyama, and M. Bolte, \href{https://doi.org/10.1063/1.3072342}{Appl. Phys. Lett. \textbf{94}, 062504 (2009)}.
\bibitem{Kamionka2010} T. Kamionka, M. Martens, K. W. Chou, M. Curcic, A. Drews, G. Sch\"{u}tz, T. Tyliszczak, H. Stoll, B. Van Waeyenberge, and G. Meier, Magnetic Antivortex-Core Reversal by Circular-Rotational Spin Currents, \href{https://doi.org/10.1103/PhysRevLett.105.137204}{Phys. Rev. Lett. \textbf{105}, 137204 (2010)}.
\bibitem{Gliga2008} S. Gliga, M. Yan, R. Hertel, and C. M. Schneider, Ultrafast dynamics of a magnetic antivortex: Micromagnetic simulations, \href{https://doi.org/10.1103/PhysRevB.77.060404}{Phys. Rev. B \textbf{77}, 060404(R) (2008)}.
\bibitem{Xing2008} X. J. Xing, Y. P. Yu, S. X. Wu, L. M. Xu, and S. W. Li, Bloch-point-mediated magnetic antivortex core reversal triggered by sudden excitation of a suprathreshold spin-polarized current, \href{https://doi.org/10.1063/1.3033400}{Appl. Phys. Lett. \textbf{93}, 202507 (2008)}.
\bibitem{Kamionka2011} T. Kamionka, M. Martens, K. W. Chou, A. Drews, T. Tyliszczak, H. Stoll, B. Van Waeyenberge, and G. Meier, Magnetic antivortex-core reversal by rotating magnetic fields, \href{https://doi.org/10.1103/PhysRevB.83.224422}{Phys. Rev. B \textbf{83}, 224422 (2011)}.
\bibitem{Pues2012} M. Pues, M. Martens, T. Kamionka, and G. Meier, Reliable nucleation of isolated magnetic antivortices, \href{https://doi.org/10.1063/1.3698150}{Appl. Phys. Lett. \textbf{100}, 162404 (2012)}.
\bibitem{Mironov2010} V. L. Mironov, O. L. Ermolaeva, S. A. Gusev, A. Yu. Klimov, V. V. Rogov, B. A. Gribkov, O. G. Udalov, A. A. Fraerman, R. Marsh, C. Checkley, R. Shaikhaidarov, and V. T. Petrashov, Antivortex state in crosslike nanomagnets, \href{https://doi.org/10.1103/PhysRevB.81.094436}{Phys. Rev. B \textbf{81}, 094436 (2010)}.
\bibitem{Haldar2013} A. Haldar and K. S. Buchanan, Magnetic antivortex formation in pound-key-like nanostructures, \href{https://doi.org/10.1063/1.4795521}{Appl. Phys. Lett. \textbf{102}, 112401 (2013)}.
\bibitem{Dzyaloshinsky1958} I. Dzyaloshinsky, A thermodynamic theory of ``weak" ferromagnetism of antiferromagnetics, \href{https://doi.org/10.1016/0022-3697(58)90076-3}{J. Phys. Chem. Solids \textbf{4}, 241 (1958)}.
\bibitem{Moriya1960} T. Moriya, Anisotropic Superexchange Interaction and Weak Ferromagnetism, \href{https://doi.org/10.1103/PhysRev.120.91}{Phys. Rev. \textbf{120}, 91  (1960)}.
\bibitem{Muhlbauer2009} S. M\"{u}hlbauer, B. Binz, F. Jonietz, C. Pfleiderer, A. Rosch, A. Neubauer, R. Georgii, and P. B\"{o}ni, Skyrmion Lattice in a Chiral Magnet, \href{https://doi.org/10.1126/science.1166767}{Science \textbf{323}, 915 (2009)}.
\bibitem{Yu2010} X. Z. Yu, Y. Onose, N. Kanazawa, J. H. Park, J. H. Han, Y. Matsui, N. Nagaosa, and Y. Tokura, Real-space observation of a two-dimensional skyrmion crystal, \href{https://doi.org/10.1038/nature09124}{Nature \textbf{465}, 901 (2010)}.
\bibitem{Heinze2011} S. Heinze, K. von Bergmann, M. Menzel, J. Brede, A. Kubetzka, R. Wiesendanger, G. Bihlmayer, and S. Bl\"{u}gel, Spontaneous atomic-scale magnetic skyrmion lattice in two dimensions, \href{https://doi.org/10.1038/nphys2045}{Nat. Phys. \textbf{7}, 713 (2011)}.
\bibitem{Sampaio2013} J. Sampaio, V. Cros, S. Rohart, A. Thiaville, and A. Fert, Nucleation, stability and current-induced motion of isolated magnetic skyrmions in nanostructures, \href{https://doi.org/10.1038/nnano.2013.210}{Nat. Nanotechnol. \textbf{8}, 839 (2013)}.
\bibitem{Hoffmann2017} M. Hoffmann, B. Zimmermann, G. P. M\"{u}ller, D. Sch\"{u}rhoff, N. S. Kiselev, C. Melcher, and S. Bl\"{u}gel, Antiskyrmions stabilized at interfaces by anisotropic Dzyaloshinskii-Moriya interactions, \href{https://doi.org/10.1038/s41467-017-00313-0}{Nat. Commun. \textbf{8}, 308 (2017)}.
\bibitem{Huang2017} S. Huang, C. Zhou, G. Chen, H. Shen, A. K. Schmid, K. Liu, and Y. Wu, Stabilization and current-induced motion of antiskyrmion in the presence of anisotropic Dzyaloshinskii-Moriya interaction, \href{https://doi.org/10.1103/PhysRevB.96.144412}{Phys. Rev. B \textbf{96}, 144412 (2017)}.
\bibitem{Camosi2018} L. Camosi, N. Rougemaille, O. Fruchart, J. Vogel, and S. Rohart, Micromagnetics of antiskyrmions in ultrathin films, \href{https://doi.org/10.1103/PhysRevB.97.134404}{Phys. Rev. B \textbf{97}, 134404 (2018)}.
\bibitem{Butenko2009} A. B. Butenko, A. A. Leonov, A. N. Bogdanov, and U. K. R\"{o}{\ss}ler, Theory of vortex states in magnetic nanodisks with induced Dzyaloshinskii-Moriya interactions, \href{https://doi.org/10.1103/PhysRevB.80.134410}{Phys. Rev. B \textbf{80}, 134410 (2009)}.
\bibitem{Luo2014} Y. M. Luo, C. Zhou, C. Won, and Y. Z. Wu, Effect of Dzyaloshinskii-Moriya interaction on magnetic vortex, \href{https://doi.org/10.1063/1.4874135}{AIP Adv. \textbf{4}, 047136 (2014)}.
\bibitem{Wang2015} J. Wang and G.-P. Li, Effect of Dzyaloshinsky-Moriya interaction on Magnetic Vortex: A real-space phase-field study, \href{https://doi.org/10.1016/j.commatsci.2015.04.010}{Comput. Mater. Sci. \textbf{108}, 316 (2015)}.
\bibitem{Luo2015} Y. M. Luo, C. Zhou, C. Won, and Y. Z. Wu, Magnetic vortex gyration affected by Dzyaloshinskii-Moriya interaction, \href{https://doi.org/10.1063/1.4919423}{J. Appl. Phys. \textbf{117}, 163916 (2015)}.
\bibitem{Liu2016} Y. Liu, M. Jia, H. Li, and A. Du, Energy analysis of a gyrating vortex with Dzyaloshinskii-Moriya interactions, \href{https://doi.org/10.1016/j.jmmm.2015.10.136}{J. Magn. Magn. Mater. \textbf{401}, 806 (2016)}.
\bibitem{Wei2023} M. Wei, Y. Hu, C. Wu, Y. Sui, and H. Li, Feature of vortex core gyration affected by Dzyaloshinskii-Moriya interaction, \href{https://doi.org/10.1016/j.cap.2022.11.009}{Curr. Appl. Phys. \textbf{46}, 8 (2023)}.
\bibitem{Mruczkiewicz2017} M. Mruczkiewicz, M. Krawczyk, and K. Y. Guslienko, Spin excitation spectrum in a magnetic nanodot with continuous transitions between the vortex, Bloch-type skyrmion, and N\'{e}el-type skyrmion states, \href{https://doi.org/10.1103/PhysRevB.95.094414}{Phys. Rev. B \textbf{95}, 094414 (2017)}.
\bibitem{Mruczkiewicz2018} M. Mruczkiewicz, P. Gruszecki, M. Krawczyk, and K. Y. Guslienko, Azimuthal spin-wave excitations in magnetic nanodots over the soliton background: Vortex, Bloch, and N\'{e}el-like skyrmions, \href{https://doi.org/10.1103/PhysRevB.97.064418}{Phys. Rev. B \textbf{97}, 064418 (2018)}.
\bibitem{Flores2020} C. Quispe Flores, C. Chalifour, J. Davidson, K. L. Livesey, and K. S. Buchanan, Semianalytical approach to calculating the dynamic modes of magnetic vortices with Dzyaloshinskii-Moriya interactions, \href{https://doi.org/10.1103/PhysRevB.102.024439}{Phys. Rev. B \textbf{102}, 024439 (2020)}.
\bibitem{Wang2020} Z. Wang, Y. Cao, R. Wang, B. Liu, H. Meng, and P. Yan, Effect of Dzyaloshinskii-Moriya interaction on magnetic vortex switching driven by radial spin waves, \href{https://doi.org/10.1016/j.jmmm.2020.167014}{J. Magn. Magn. Mater. \textbf{512}, 167014 (2020)}.
\bibitem{Li2022} H.-N. Li, T.-X. Xue, L. Chen, Y.-R. Sui, and M.-B. Wei, Influence of Dzyaloshinskii-Moriya interaction on the magnetic vortex reversal in an off-centered nanocontact geometry, \href{https://doi.org/10.1088/1674-1056/ac4cbd}{Chin. Phys. B \textbf{31}, 097501 (2022)}.
\bibitem{Siracusano2016} G. Siracusano, R. Tomasello, A. Giordano, V. Puliafito, B. Azzerboni, O. Ozatay, M. Carpentieri, and G. Finocchio, Magnetic Radial Vortex Stabilization and Efficient Manipulation Driven by the Dzyaloshinskii-Moriya Interaction and Spin-Transfer Torque, \href{https://doi.org/10.1103/PhysRevLett.117.087204}{Phys. Rev. Lett. \textbf{117}, 087204 (2016)}.
\bibitem{Karakas2018} V. Karakas, A. Gokce, A. T. Habiboglu, S. Arpaci, K. Ozbozduman, I. Cinar, C. Yanik, R. Tomasello, S. Tacchi, G. Siracusano, M. Carpentieri, G. Finocchio, T. Hauet, and O. Ozatay, Observation of Magnetic Radial Vortex Nucleation in a Multilayer Stack with Tunable Anisotropy, \href{https://doi.org/10.1038/s41598-018-25392-x}{Sci. Rep. \textbf{8}, 7180 (2018).}.
\bibitem{Rohart2013} S. Rohart and A. Thiaville, Skyrmion confinement in ultrathin film nanostructures in the presence of Dzyaloshinskii-Moriya interaction, \href{https://doi.org/10.1103/PhysRevB.88.184422}{Phys.  Rev. B \textbf{88}, 184422 (2013)}.
\bibitem{Vansteenkiste2014} A. Vansteenkiste, J. Leliaert, M. Dvornik, M. Helsen, F. Garcia-Sanchez, and B. Van Waeyenberge, The design and verification of MuMax3, \href{https://doi.org/10.1063/1.4899186}{AIP Adv. \textbf{4}, 107133 (2014)}.
\bibitem{Metlov2002} K. L. Metlov and K. Y. Guslienko, Stability of magnetic vortex in soft magnetic nano-sized circular cylinder, \href{https://doi.org/10.1016/S0304-8853(01)01360-9}{J. Magn. Magn. Matter. \textbf{242}, 1015 (2002)}.
\bibitem{Metlov2008} K. L. Metlov and Y. P. Lee, Map of metastable states for thin circular magnetic nanocylinders, \href{https://doi.org/10.1063/1.2898888}{Appl. Phys. Lett. \textbf{92}, 112506 (2008)}.
\bibitem{Verba2018} R. V. Verba, D. Navas, A. Hierro-Rodriguez, S. A. Bunyaev, B. A. Ivanov, K. Y. Guslienko, and G. N. Kakazei, Overcoming the Limits of Vortex Formation in Magnetic Nanodots by Coupling to Antidot Matrix, \href{https://doi.org/10.1103/PhysRevApplied.10.031002}{Phys. Rev. Applied \textbf{10}, 031002 (2018)}.
\bibitem{Verba2020} R. V. Verba, D. Navas, S. A. Bunyaev, A. Hierro-Rodriguez, K. Y. Guslienko, B. A. Ivanov, and G. N. Kakazei, Helicity of magnetic vortices and skyrmions in soft ferromagnetic nanodots and films biased by stray radial fields, \href{https://doi.org/10.1103/PhysRevB.101.064429}{Phys.  Rev. B \textbf{101}, 064429 (2020)}.
\bibitem{Li2020} C. Li, S. Wang, N. Xu, X. Yang, B. Liu, B. Yang, and L. Fang, Spin-Torque Nano-Oscillators Based on Radial Vortex in the Presence of Interface Dzyaloshinskii-Moriya Interaction, \href{https://doi.org/10.1016/j.jmmm.2019.166155}{J. Magn. Magn. Mater. \textbf{498}, 166155 (2020)}.
\bibitem{Lee2011} K.-S. Lee, M.-W. Yoo, Y.-S. Choi, and S.-K. Kim, Edge-Soliton-Mediated Vortex-Core Reversal Dynamics, \href{http://dx.doi.org/10.1103/PhysRevLett.106.147201}{Phys. Rev. Lett. \textbf{106}, 147201 (2011)}.
\bibitem{Waeyenberge2006} B. Van Waeyenberge, A. Puzic, H. Stoll, K. W. Chou, T. Tyliszczak, R. Hertel, M. F\"{a}hnle, H. Br\"{u}ckl, K. Rott, G. Reiss, I. Neudecker, D. Weiss, C. H. Back and G. Sch\"{u}tz, Magnetic Vortex Core Reversal by Excitation with Short Bursts of an Alternating Field, \href{https://doi.org/10.1038/nature05240}{Nature \textbf{444}, 461 (2006)}.
\bibitem{Ma2019} Y. Ma, R. Zhao, C. Song, C. Jin, J. Wang, Y. Wei, Y. Huang, J. Wang, J. Wang, and Q. Liu, Current-driven radial vortex switching in a permalloy nanodisk, \href{https://doi.org/10.1016/j.jmmm.2019.165544}{J. Magn. Magn. Mater. \textbf{491}, 165544 (2019)}.
\bibitem{Curcic2008} M. Curcic, B. Van Waeyenberge, A. Vansteenkiste, M. Weigand, V. Sackmann, H. Stoll, M. F\"{a}hnle, T. Tyliszczak, G. Woltersdorf, C. H. Back, and G. Sch\"{u}tz, Polarization Selective Magnetic Vortex Dynamics and Core Reversal in Rotating Magnetic Fields, \href{https://doi.org/10.1103/PhysRevLett.101.197204}{Phys. Rev. Lett. \textbf{101}, 197204 (2008)}.
\bibitem{Yu2020} D. Yu, J. Kang, J. Berakdar, and C. Jia, Nondestructive ultrafast steering of a magnetic vortex by terahertz pulses, \href{https://doi.org/10.1038/s41427-020-0217-8}{NPG Asia Mater. \textbf{12}, 36 (2020)}.
\bibitem{Yu2021} D. Yu, C. Sui, D. Schulz, J. Berakdar, and C. Jia, Nanoscale Near-Field Steering of Magnetic Vortices, \href{http://dx.doi.org/10.1103/PhysRevApplied.16.034032}{Phys. Rev. Applied \textbf{16}, 034032 (2021)}.
\bibitem{Ostler2015} T. A. Ostler, R. Cuadrado, R. W. Chantrell, A. W. Rushforth, and S. A. Cavill, Strain Induced Vortex Core Switching in Planar Magnetostrictive Nanostructures, \href{http://dx.doi.org/10.1103/PhysRevLett.115.067202}{Phys. Rev. Lett. \textbf{115}, 067202 (2015)}.
\bibitem{Zhu2021} M. Zhu, H. Hu, S. Cui, Y. Li, X. Zhou, Y. Qiu, R. Guo, G. Wu, G. Yu, and H. Zhou, Strain-driven radial vortex core reversal in geometric confined multiferroic heterostructures, \href{https://doi.org/10.1063/5.0054010}{Appl. Phys. Lett. \textbf{118}, 262412 (2021)}.
\bibitem{Crepieux1998} A. Cr\'{e}pieux and C. Lacroix, Dzyaloshinsky-Moriya interactions induced by symmetry breaking at a surface, \href{https://doi.org/10.1016/S0304-8853(97)01044-5}{J. Magn. Magn. Mater. \textbf{182}, 341 (1998)}.
\bibitem{Camosi2017} L. Camosi, S. Rohart, O. Fruchart, S. Pizzini, M. Belmeguenai, Y. Roussign\'{e}, A. Stashkevich, S. M. Cherif, L. Ranno, M. de Santis, and J. Vogel, Anisotropic Dzyaloshinskii-Moriya interaction in ultrathin epitaxial Au/Co/W(110), \href{https://doi.org/10.1103/PhysRevB.95.214422}{Phys. Rev. B \textbf{95}, 214422 (2017)}.
\bibitem{Nayak2017} A. K. Nayak, V. Kumar, T. Ma, P. Werner, E. Pippel, R. Sahoo, F. Damay, U. K. R\"{o}{\ss}ler, C. Felser, and S. S. P. Parkin, Magnetic antiskyrmions above room temperature in tetragonal Heusler materials, \href{https://doi.org/10.1038/nature23466}{Nature \textbf{548}, 561 (2017)}.
\bibitem{Sun2023} J. Sun, S. Shi, P. Han, Y. Wang, Y. Zhao, B.-X. Xu, and J. Wang, Strain Mediated Transition between Skyrmion and Antiskyrmion in Ferromagnetic Thin Films, \href{http://dx.doi.org/10.2139/ssrn.4674228}{ssrn.4674228}.
\bibitem{Cheng2024} Z. Cheng, H. Chen, S. Huang, and Y. Wu, Spin wave mode coupling in antiskyrmions induced by rotational symmetry breaking, \href{https://doi.org/10.1103/PhysRevB.109.104431}{Phys. Rev. B \textbf{109}, 104431 (2024)}.
\bibitem{Kim2012} J.-H. Kim, K.-S. Lee, H. Jung, D.-S. Han, and S.-K. Kim, Information-signal-transfer rate and energy loss in coupled vortex-state networks, \href{http://dx.doi.org/10.1063/1.4748885}{Appl. Phys. Lett. \textbf{101}, 092403 (2012)}.
\bibitem{Han2013} D.-S. Han, A. Vogel, H. Jung, K.-S. Lee, M. Weigand, H. Stoll, G. Sch\"{u}tz, P. Fischer, G. Meier, and S.-K. Kim, Wave modes of collective vortex gyration in dipolar-coupled-dot-array magnonic crystals, \href{https://doi.org/10.1038/srep02262}{Sci. Rep. \textbf{3}, 2262 (2013)}.
\bibitem{Li2019} Z.-X. Li, Y. Cao, P. Yan, and X. R. Wang, Higher-order topological solitonic insulators, \href{https://doi.org/10.1038/s41524-019-0246-4}{NPJ Comput. Mater. \textbf{5}, 107 (2019)}.
\bibitem{Li202002} Z.-X. Li, Y. Cao, X. R. Wang, and P. Yan, Symmetry-Protected Zero Modes in Metamaterials Based on Topological Spin Texture, \href{https://doi.org/10.1103/PhysRevApplied.13.064058}{Phys. Rev. Applied \textbf{13}, 064058 (2020)}.
\bibitem{Ma2020} Y. Ma, C. Song, C. Jin, Z. Zhu, H. Feng, H. Xia, J. Wang, Z. Zeng, J. Wang, and Q. Liu, Nano-oscillator based on radial vortex by overcoming the switching of core, \href{https://doi.org/10.1088/1361-6463/ab71af}{J. Phys. D: Appl. Phys. \textbf{53}, 195004 (2020)}.
\bibitem{Garcia2016} F. Garcia-Sanchez, J. Sampaio, N. Reyren, V. Cros, and J.-V. Kim, A skyrmion-based spin-torque nano-oscillator, \href{http://dx.doi.org/10.1088/1367-2630/18/7/075011}{New J. Phys. \textbf{18}, 075011 (2016)}.

\end{thebibliography}
\end{document}